\documentclass[prl,twocolumn,showpacs,floatfix]{revtex4-1}
\usepackage{epsfig,subfigure,amssymb,amscd,amsmath,graphicx,amsfonts}
\usepackage{times}

\newcommand{\non}{\nonumber}

\newcommand{\bea}{\begin{eqnarray}}
\newcommand{\eea}{\end{eqnarray}}
\newcommand{\be}{\begin{equation}}
\newcommand{\ee}{\end{equation}}
\newcommand{\ba}{\begin{align}}
\newcommand{\ea}{\end{align}}

\newcommand{\ket}[1]{     |    \,    #1    \rangle}
 
\newcommand{\ZZ}{\mathbb{Z}}

\newcommand{\bk}{{\boldsymbol{k}}}

\newcommand{\lm}{\lambda_-}
\newcommand{\lp}{\lambda_+}
\usepackage[usenames]{color}

\begin{document}

\title{Low-energy sub-gap states in multi-channel Majorana wires}

\author{G. Kells, D. Meidan and P. W. Brouwer}

\affiliation{Dahlem Center for Complex Quantum Systems and Fachbereich Physik, Freie Universit\"{a}t Berlin, Arnimallee 14, 14195 Berlin, Germany}

\begin{abstract}
One-dimensional p-wave superconductors are known to harbor Majorana bound states at their ends. Superconducting wires with a finite width $W$ may have fermionic subgap states in addition to possible Majorana end states. 
While they do not necessarily inhibit the use of Majorana end states for topological computation, these subgap states can obscure the identification of a topological phase through a density-of-states measurement. We present two simple models to describe low-energy fermionic subgap states. If the wire's width $W$ is much smaller than the superconductor coherence length $\xi$, the relevant subgap states are localized near the ends of the wire and cluster near zero energy, whereas the lowest-energy subgap states are delocalized if $W \gtrsim \xi$. Notably, the energy of the lowest-lying fermionic subgap state (if present at all) has a maximum for $W \sim \xi$.
\end{abstract}


\date{\today} \maketitle

The search for Majorana fermions has attracted a great deal of interest in the last few years \cite{kn:wilczek2009}. Notably their non-local properties and non-abelian braiding statistics make Majorana fermion systems attractive candidates for fault tolerant quantum computation \cite{kn:freedman1998,kn:kitaev2003,Nayak2008}. The present wave of interest is driven by a number of proposals that suggest ways of realizing and manipulating Majorana states in solid state systems, most prominently interfaces of s-wave superconductors and topological insulators \cite{Fu2008,Cook2011}, half-metallic ferromagnets \cite{Duckheim2011,kn:chung2011,kn:weng2011}, or semiconductor films or wires \cite{Oreg2010,Lutchyn2010,Lutchyn2011}, where the latter stand out because Majorana manipulation require a mere series of gate operations \cite{kn:alicea2011}. In all these proposals, the proximity coupling to the s-wave superconductor effectively turns the normal metal into a p-wave superconductor, which is well known to harbor Majorana fermions at its ends or edges \cite{Jackiw1976,Kitaev2001,Read2000,Motrunich2001}.

Majorana bound states at ends of what is effectively a p-wave superconducting wire can be analyzed most straightforwardly if these wires are strictly one dimensional, with only a single propagating mode at the Fermi level in the absence of superconductivity \cite{Oreg2010,Lutchyn2010}. Nevertheless, Majorana end states can also exist in a quasi one-dimensional geometry. The effect of multiple transverse channels, present in most realistic realizations, has been addressed in Refs.\ \cite{Wimmer2010,Potter2010,Lutchyn2011,Stanescu2011,Zhou2011}. Specifically one sees that a complex p-wave superconductor in a strip geometry  
undergoes a series of oscillatory quantum phase transitions between topologically trivial and topologically non-trivial phases (without and with Majorana end states, respectively) as the strip width $W$ or chemical potential $\mu$ are varied. Both with and without Majorana end states, a range of sub-gap states is found \cite{Potter2010}, analogous to the sub-gap states in vortex cores of bulk superconductors \cite{Caroli1964}. Although the mere presence of sub-gap states does not prohibit the use of Majorana end states for topological quantum computation \cite{Akhmerov2010}, the presence of low-lying sub-gap states clearly obstructs an unambiguous experimental verification of the Majorana states.

\begin{figure}
\centering
\includegraphics[width=.46\textwidth,height=0.25\textwidth]{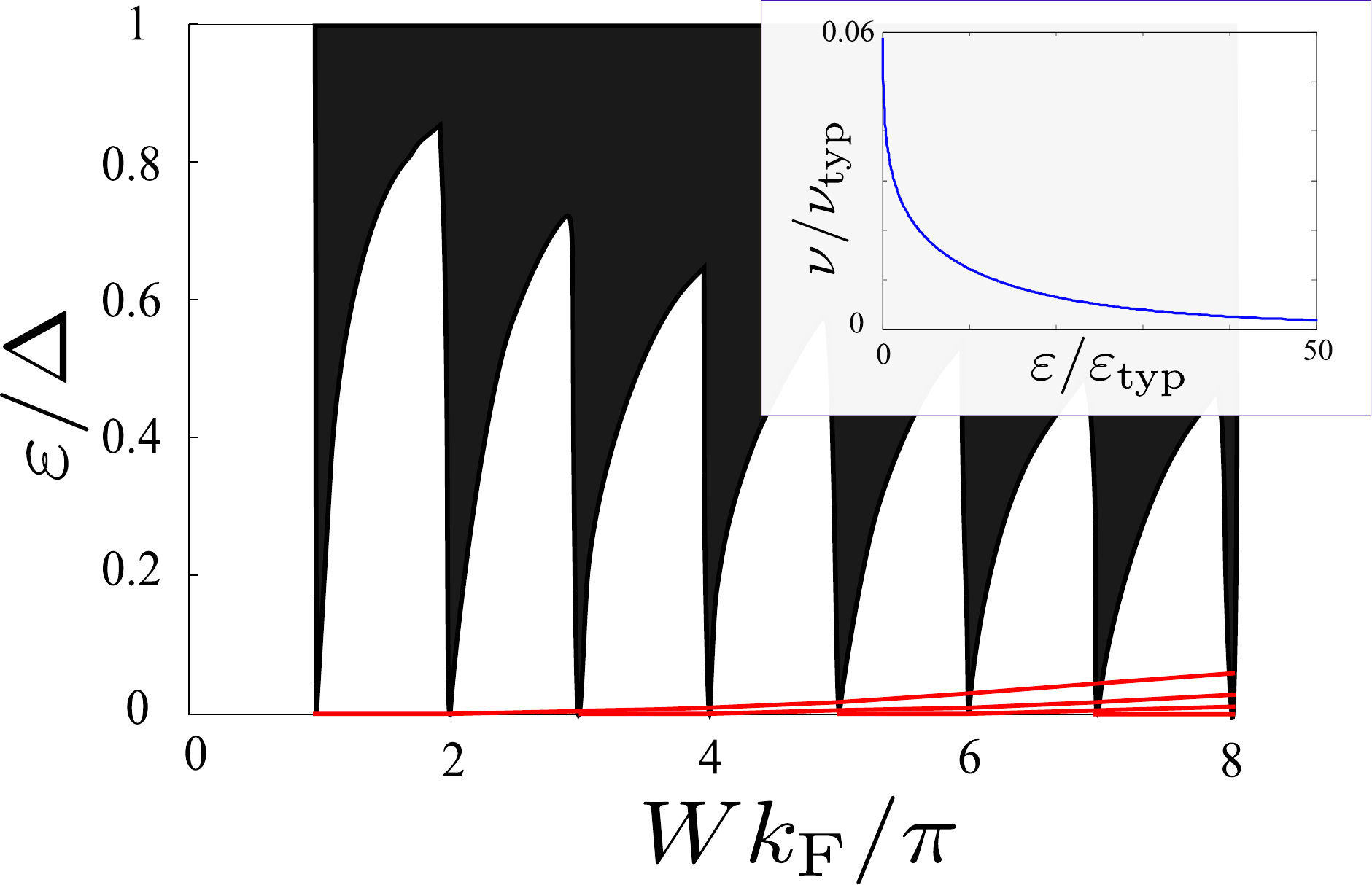}
\caption[]{\label{fig:Gap4} (Color online) Excitation spectrum of a p-wave superconducting strip as a function of the strip width $W$, for small $Wk_F$. The solid (red) curves and the shaded region indicate the discrete energies of the end states and the continuous spectrum, respectively. The data shown in the figure is obtained from a solution of the effective models described in the text with $k_{\rm F}\xi = 100$. Data points obtained from a numerical solution of the lattice with  $-4 t < \mu < -3t $ are indistinguishable from the curves shown here. Inset: Density of fermionic end states in the limit $\lambda_{\rm F} \ll W \ll \xi$. Here $\epsilon_\textrm{typ} = \pi \Delta (W/\xi)^2 $ and $\nu_\textrm{typ} = Wk_F/(\pi \epsilon_\textrm{typ}). $}
\end{figure}

\begin{figure}
\centering
\includegraphics[width=.44\textwidth,height=0.22\textwidth]{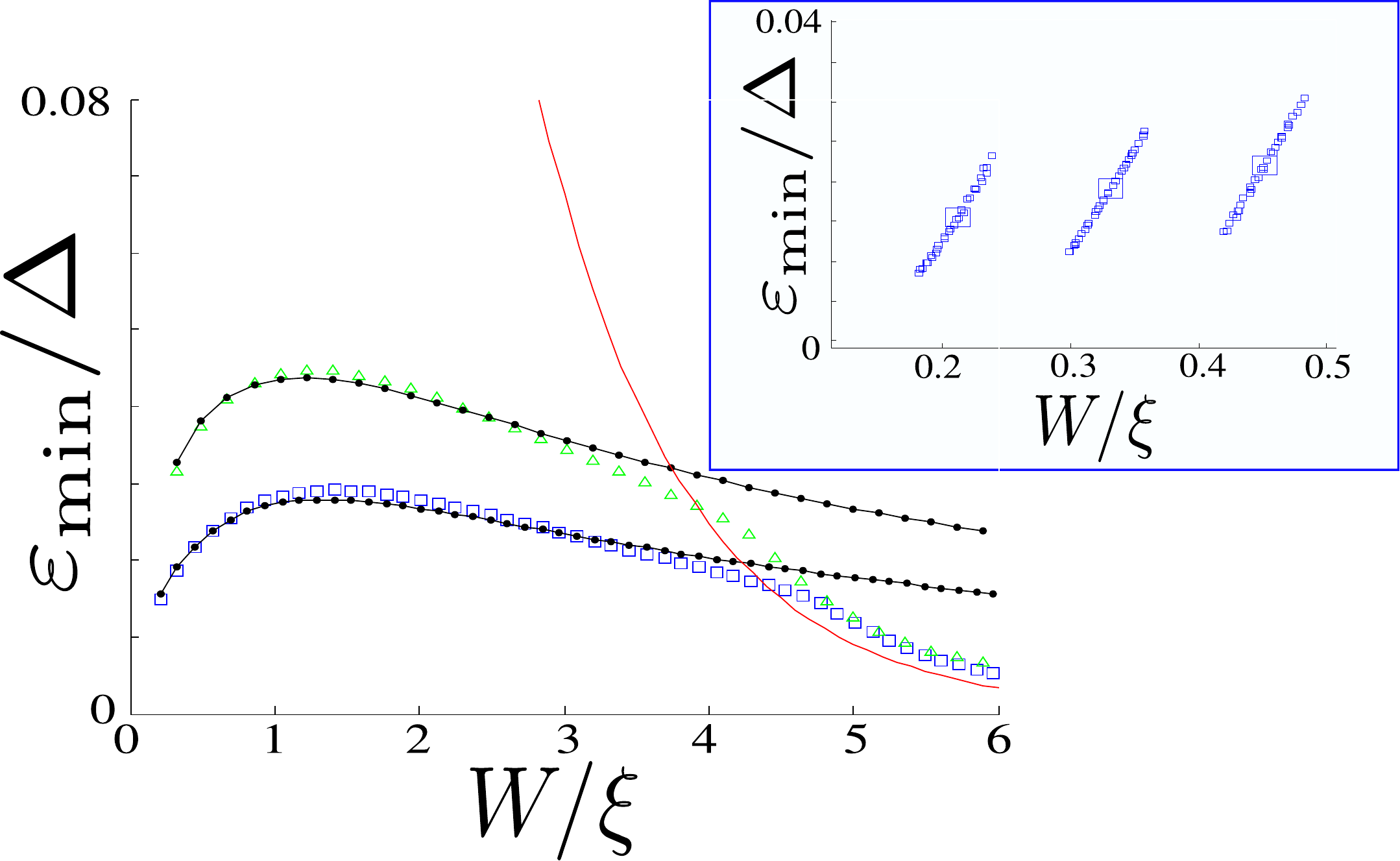}
\caption[]{\label{fig:Gap3} (Color online) The energy $\varepsilon_{\rm min}$ of the lowest-lying fermionic excitation in the topological non trivial phases, as a function of the sample width $W$. The main panel shows the value of $\varepsilon_{\rm min}$  at the mid-point of each topological non-trivial phase for $\mu=-3t$ (see inset), which illustrates the large-scale dependence of $\varepsilon_{\rm min}$ on the sample width $W$. Squares and triangles refer to full numerical solutions of the lattice model (\ref{eq:Hp}) for $k_F \xi=50$ and $33\frac{1}{3}$ respectively; The dotted solid black curves and the red solid curve represent solutions of the effective models (\ref{eq:Hmn}) and (\ref{eq:Hkx}), respectively. Inset: $\varepsilon_{\text{min}}$ vs. $W$ for the topologically non-trivial phases for $k_F \xi=50$, obtained by combining data from a numerical solution of the lattice model (\ref{eq:Hp}) for various values of $\mu$ near $-3t$.} 
\end{figure}

The purpose of this paper is to systematically analyse the energies of the sub-gap states as a function of the sample width $W$. Throughout we assume that $\xi \gtrsim \lambda_{\rm F}$, where $\lambda_{\rm F} = 2 \pi/k_{\rm F}$ is the Fermi wavelength in the absence of superconductivity and $\xi$ the superconducting coherence length. Our results, which will be discussed in detail below, reveal  
two different regimes, depending on the relative magnitude of $W$ and $\xi$:
\begin{enumerate}
\item If $W \lesssim \xi$ there is an alternation of topological and non-topological phases, where each phase harbors at most one Majorana state at zero energy and $N_{\rm f} \le N= \mbox{int}\, [(W/\pi)\sqrt{k_{\rm F}^{2} - \xi^{-2}}]$ fermionic subgap states localized near each end of the wire, see Fig.\ \ref{fig:Gap4}. The lowest-lying fermionic subgap state (if present) has energy 
\begin{equation}
  \varepsilon_{\rm min} \sim \Delta \frac{W \lambda_{\rm F}}{\xi^2 \ln(W/\lambda_{\rm F})}.
  \label{eq:Wsmall}
\end{equation}
For wire widths $W \ll \xi^{2/3} \lambda_{\rm F}^{1/3}$ one has $N_{\rm f} = N$ and the highest-lying subgap state has energy $\sim \Delta W^3/\xi^2 \lambda_{\rm F}$, well below the bulk excitation gap $\Delta$.
\item For $W \gtrsim \xi$ there is an alternation of topological and non-topological phases, with Majorana end states in the topological phase, and delocalized subgap states with minimum energy 
\begin{equation}
  \varepsilon_{\rm min} \sim \Delta e^{-W/\xi}.
  \label{eq:Wlarge}
\end{equation}
Equation (\ref{eq:Wlarge}) was obtained previously in Ref.\ \onlinecite{Zhou2011}.
\end{enumerate}
We note that as most experimental proposals involve a rather weak induced superconductivity because of the presence of Schottky barrier or the dependence on spin-orbit coupling, we believe that the quasi one-dimensional regime $W/\xi \ll 1$ is most relevant for possible applications.
Hence, these finding imply that, in the experimentally accessible regime, localized subgap states cluster around zero energy, thus obstructing the experimental verification of Majorana states through a density-of-states measurement, see Fig.~\ref{fig:Gap4}. In fact, the optimal wire width $W$ that gives rise to the largest excitation gap to fermionic subgap states is $W \sim \xi$ (Fig. \ref{fig:Gap3}), although the optimal excitation gap for a multi-channel wire is always lower than that of a single channel wire. Nonetheless, for  $W/\xi \ll 1$ there is no coupling between subgap states localized at opposite ends of the strip, as the excitation gap to delocalized bulk states is large ($\gtrsim \Delta $, see Fig.~\ref{fig:Gap4}), making the narrow limit favorable  for topological protection.   Our analysis provides a straightforward physical interpretation of the oscillatory quantum phase transitions and highlights broad similarities between the $\ZZ$ to $\ZZ_2$ crossover and the nucleation of topological phases discussed for example in \cite{Gils2009,Ludwig2011}.

For the detailed discussion of these findings we employ a two-dimensional lattice model of a complex p-wave superconductor \cite{Wilson1974,Kitaev2001,Motrunich2001}. The same model was studied recently in Refs.\ \cite{Wimmer2010,Potter2010,Lee2007,Bray-Ali2009}. We consider a strip geometry of length $ L=N_x a$ and width $W = N_y a $, where $a$ is the lattice constant, with the Hamiltonian 
\begin{eqnarray}
  H &=& 
  - \mu \sum_{i=1}^{N_x} \sum_{j=1}^{N_y} c^{\dagger}_{i,j} c_{i,j} 
  \nonumber \\ && \mbox{}
  - \sum_{i=1}^{N_x-1} \sum_{j=1}^{N_y} 
  \left( 
  t c^{\dagger}_{i+1,j} c_{i,j} 
  - i \Delta_x c^{\dagger}_{i+1,j} c^{\dagger}_{i,j} + \mbox{h.c.} 
  \right) \nonumber \\ && \mbox{}
  - \sum_{i=1}^{N_x} \sum_{j=1}^{N_y-1} 
  \left(t c^{\dagger}_{i,j+1} c_{i,j}  
  + \Delta_y c^{\dagger}_{i,j+1} c^{\dagger}_{i,j} + \mbox{h.c.} 
  \right),~~~
 \label{eq:Hp} 
\end{eqnarray}
where $\mu$ is the chemical potential, $t$ the hopping amplitude, and $\Delta_x$ and $\Delta_y$ set the p-wave pairing amplitudes in the longitudinal and transverse directions, respectively. We will focus on the scenario that the p-wave superconducting order is inherited from proximity coupling to a bulk superconductor, as in the proposals of Refs.\ \onlinecite{Fu2008,Cook2011,Duckheim2011,kn:chung2011,kn:weng2011,Oreg2010,Lutchyn2010,Lutchyn2011}, so that no self-consistency conditions need to be accounted for. The bulk dispersion of Eq.\ (\ref{eq:Hp}) gives excitations at energies
\begin{eqnarray}
 E_{\bk}^2 &=& [-\mu - 2 t(\cos k_x a + \cos k_y a)]^2
  \nonumber \\ && \mbox{} + 4\Delta_x^2 \sin^2 k_x a + 4\Delta_y^2 \sin^2 k_y a.
\end{eqnarray}
Below we will set $\Delta_x = \Delta_y$ and formulate our main results in the continuum limit, which is obtained by sending $a \to 0$ while keeping $L$, $W$, and the number of electrons fixed. In this limit, $\mu \to -4 t + t (k_{\rm F} a)^2$ and $E_k \to k_{\rm F}^{-1} \sqrt{(k^2-k_{\rm F}^2)^2 (\hbar v_{\rm F})^2 + k^2 \Delta^2}$, where $k_{\rm F}= 2\pi/\lambda_{\rm F}$ is the Fermi wavenumber, $\hbar v_{\rm F} = 2 t k_{\rm F} a^2$ the Fermi velocity, and $\Delta = 2 \Delta_x k_{\rm F} a = 2 \Delta_y k_{\rm F} a$ the bulk excitation gap. The superconductor coherence length is $\xi = \hbar v_{\rm F}/\Delta$. 
The model (\ref{eq:Hp}) falls into class $D$ in the symmetry classification of non-interacting topological insulators and superconductors \cite{Schnyder2008,Kitaev2009}. In one dimension, the archetypal model in the $D$ class is the so called Majorana chain or wire \cite{Kitaev2001}, which is characterized by a $\ZZ_2$ topological invariant. In the two dimensional (2d) limit, class $D$  is characterized by a $\ZZ$ topological invariant. This model, and its $k_x - ik_y$ partner correspond to realizations with Chern numbers of $\sigma = 0 ,\pm1$; examples of  higher Chern numbers do exist, see {\em e.g.}, \cite{Read2000,Lahtinen2010,Mao2011,Kells2010c}.

For $W < \pi(k_{\rm F}^2 - \xi^2)^{-1/2}$ the model (\ref{eq:Hp}) is a trivial insulator. Upon increasing the sample width, the system undergoes a series of topological phase transitions. In order to study low-lying excitations in each phase, we use two separate calculations, valid for $W \ll \xi$ and $W \gtrsim \xi$, respectively.

{\em The case $W \ll \xi$.} If $W \ll \xi$, hypothetical end states have a vanishing expectation value of the transverse momentum $p_y$. For this reason, it is a good approximation to treat the term proportional to $\Delta_y$ in the Hamitonian (\ref{eq:Hp}) in perturbation theory. Solving for the eigenstates of Eq.\ (\ref{eq:Hp}) with $\Delta_y=0$ in the limit of large $L$, we find (in the continuum limit) 
\begin{equation}
  N = \mbox{int}\, [(W/\pi)\sqrt{k_{\rm F}^{2} - \xi^{-2}}]
  \label{eq:N}
\end{equation}
zero-energy Majorana states at the left ($+$) or right ($-$) ends of the wire, with creation/annihilation operators $\gamma_{n,\pm} = e^{\mp i \pi/4} \sum_{i,j} u_{n,\pm}(ia,ja) (c_{i,j} \pm i c_{i,j}^{\dagger})$, $n=1,2,\ldots,N$, and
\begin{equation}
  u_{n,+}(x,y) = \sqrt{\frac{2}{W}}\Omega_n^{-1/2}  \sin(n \pi y/W)
  \sin(k_{nx} x)e^{-x/\xi} .\ \
\label{eq:psin}
\end{equation}
Here $k_{nx}^2 = k_{\rm F}^2 - (n \pi/W)^2 - \xi^{-2}$, $ \Omega_n$ is a normalization constant, and  $u_{n,-}(x,y) = u_{n,+}(L-x,y)$. Full results for the lattice model are qualitatively similar \footnote{See supporting material for details.}. The pairing term proportional to $\Delta_y$ is then treated in degenerate perturbation theory. Using the $N$ zero-energy states of Eq.\ (\ref{eq:psin}) at each end as a basis, one finds the effective $N \times N$ Hamiltonian $H_{nm}$ with $H_{nm} = 0$ if $n+m$ is even and
\bea
\label{eq:Hmn}
  H_{nm} &=& 
  \frac{32 i \Delta \lambda_{\rm F} m n}
  {\pi W (m^2-n^2)}
  \\ && \mbox{} \times\nonumber 
  \frac{\sqrt{(4 W^2 - \lambda_{\rm F}^2 n^2)
  (4 W^2 - \lambda_{\rm F}^2 m^2)}}{
  (8 W)^2 -8 \lambda_{\rm F}^2 (n^2 + m^2)
  + \frac{\lambda_{\rm F}^2 \xi^2}{W^2} (n^2-m^2)^2}
\eea
if $n+m$ is odd.
The eigenvalues $\varepsilon$ of the antisymmatric matrix (\ref{eq:Hmn}) come in pairs $\pm \varepsilon$; the Hamiltonian (\ref{eq:Hmn}) has a single zero eigenvalue if $N$ is odd. In that case the corresponding zero mode at the left/right end is a linear combination of orbitals $u_{n,\pm}(x,y)$ with $n$ odd only. The topologically trivial phases, which have no zero-energy end states, occur for even $N$. Since $N$ increases stepwise as the chemical potential $\mu$ or the width $W$ is varied, the system thus undergoes a sequence of topological phase transitions. We could not obtain a closed-form expression for the eigenvalues of the effective Hamiltonian (\ref{eq:Hmn}) for general $N$. A simple scaling analysis of the Hamiltonian (\ref{eq:Hmn}) shows that  the lowest positive eigenvalue $\varepsilon_{\rm min}$ is of order $\Delta \lambda_{\rm F} \min(W/\xi^2,1/W)$, so that the optimal (largest) separation between the Majorana state and the lowest-lying fermionic excitation is achieved for $W$ of order $\xi$. A numerical evaluation of the spectrum reveals an additional logarithmic correction in the limit $W \ll \xi$, thus giving the estimate (\ref{eq:Wsmall}) for $W \ll \xi$. The highest eigenvalue of the matrix (\ref{eq:Hmn}) is of order $\Delta \max(1,W^3/\lambda_{\rm F} \xi^2)$. For $W \lesssim \xi$, the density of states has median $\varepsilon_{\text{typ}} \sim \pi \Delta (W/\xi)^2$, well below the bulk excitation gap $\Delta$, thus justifying the perturbative procedure, see Fig.\ \ref{fig:Gap4} (inset). 

For $W \sim \xi$, the spectrum of end states found here is similar to that of the subgap states in a vortex core \cite{Caroli1964}, which appear at a regular spacing $\sim \Delta \lambda_{\rm F}/\xi$, filling up the entire region between zero excitation energy and the bulk gap $\Delta$. On the other hand, in the limit that the wire width is small in comparison to the coherence length, the end states cluster near zero excitation energy and their spacing is anomalously small.

{\em The case $W \gtrsim \xi$.}
The end states discussed sofar are not the only possible low-energy excitations of the system. In addition to the end states, there are lateral Majorana modes localized near the edges at $y=0$ and $y=W$ \cite{Jackiw1976,Read2000}. 
At transition points between the topological phases, the lowest energies of these lateral edge modes drops to zero --- see the discussion below Eq.\ (\ref{eq:uWlarge}) for details ---, so that the effective Hamiltonian (\ref{eq:Hmn}) provides a complete description of the low-energy excitations away from the transition points only. Moreover, away from the transition points,  the energy splitting of the lateral modes  decreases exponentially as the width increases, and eventually they become the dominant low-energy excitation for sufficiently large $W \gtrsim \xi$, with an excitation gap that follows $\sim \Delta e^{-W/\xi}$.

In order to gain a deeper understanding of the role of the lateral Majorana modes, we now apply periodic boundary conditions in the longitudinal ($x$) direction, while keeping the hard boundaries at $y=0$ ($+$) and $y=W$ ($-$). The calculation presented here allows one to calculate the  excitation gap to the lateral edge modes in the continuum limit and reveals that the $\ZZ_2$ topological order of the bulk can be understood  as effective Majorana hopping model. The Majorana modes on the lateral edges are labeled by the wavenumber $k_x$ and have operators $\gamma_{k_x,\pm} = \sum_{m,n} u_{k_x,\pm}(na) e^{i k_x m a} (c_{m,n}  \pm c_{m,n}^{\dagger} )$ and energy $\varepsilon_{\pm}(k_{x})$. In the limit $W/\xi \to \infty$ the wavefunctions $u_{k_{x},\pm}(y)$ are known \cite{Read2000},
\begin{equation}
  u_{k_x,+}(y) = \frac{e^{-y/\xi} \sin(k_y y)}{\sqrt{\Omega_{k_x}}}, \ \
  k_y = \sqrt{k_{\rm F}^2 - k_x^2 - \xi^{-2}},
  \label{eq:uWlarge}
\end{equation}
with $\Omega_{k_x}$ a normalization constant
and $u_{k_x,-}(y) = i u_{k_x,+}(W-y)$,  and their energy is
\be
  \varepsilon_{\pm}(k_x) = \pm \xi(k_x),\ \ 
  \xi(k_x) = k_x \Delta/k_{\rm F}.
\ee
The wavefunctions (\ref{eq:uWlarge}) also satisfy the boundary conditions for finite $W$ if $k_y W/\pi$ is an integer. These special points, at which the spectrum of the lateral Majorana modes is gapless, correspond to the boundaries between the topological and non-topological phases \cite{Zhou2011}. [Note the consistency with Eq.\ (\ref{eq:N}) above.] In the vicinity of the phase boundary at $k_y = k_y^{\rm c} = m_y \pi/W$, $m_y$ integer, the energies of the lateral Majorana modes can be obtained in perturbation theory, which leads to the effective $2 \times 2$ matrix Hamiltonian
\be
  H_{k_x} = \left( \begin{array}{cc}
  \xi(k_x) & h(k_x) \\ h(k_x) & - \xi(k_x) \end{array} \right),
  \label{eq:Hkx}
\ee
for the lateral Majorana modes, with
\be
  h(k_x) = -i \frac{\hbar v_{\rm F}}{2k_{\rm F}} [(k_y^{{\rm c}})^2 - k_y^2] \langle u_{k_x,+}^{\rm c} | u_{k_x,-}^{\rm c} \rangle,
\ee
where the superscript ``c'' refers to the wavefunctions at the transition point, {\em i.e.}, with $k_y = k_y^{\rm c}$.
The eigenvalues of $H_{k_x}$ are $\varepsilon_{\pm}(k_x) = \pm [h(k_x)^2 + \xi(k_x)^2]^{1/2}$. 
Interestingly, the sign of $h(k_x)$ encodes the sign of the effective mass of the system \cite{Jackiw1976} and therefore also encodes the $\ZZ_2$ topological index of the system \cite{Kitaev2001,Read2000}.
Although the effective Hamiltonian (\ref{eq:Hkx}) has been derived in the vicinity of transition points between topological and non-topological phases ($|k_y-k_y^{\rm c}|$ small), comparison with a full numerical solution of the lattice model (\ref{eq:Hp}) shows good agreement for all $W$ if $m_y$ is taken to be the integer closest to $k_y W/\pi$. (Agreement within $13 \%$ for $W \gg \xi$, data not shown.)
Making use of the estimate $\langle u_{k_x,+} | u_{k_x,-} \rangle \approx (-1)^{m+1} (W/2 \xi) e^{-W/\xi}$ for $W \gtrsim \xi$, we find that the typical gap $\varepsilon_{\rm min}$ of the lateral Majorana modes away from the transition points is given by Eq.\ (\ref{eq:Wlarge}) if $W \gg \xi$.

Figure \ref{fig:Gap3} shows the value of the energy $\varepsilon_{\rm min}$  of the lowest-lying fermionic excitation at the mid-point of each topological phase, as a function of the width $W$. We observe that the two calculations outlined above are in excellent agreement with the full numerical calculation in the limits $W \ll \xi$ and $W \gg \xi$. Although there are quantitative deviations in the crossover between these two limits, $\varepsilon_{\rm min}$ is qualitatively well described by the minimum of the two asymptotic expressions (\ref{eq:Wlarge}) and (\ref{eq:Wsmall}).

{ \em Conclusion and perspective - } We have presented two simple analytical models to describe two types of low-lying fermionic excitations in a multichannel p wave superconducting wire.  
The first model describes fermionic end states, which are the lowest relevant excitations for wire widths $W$ well below the superconductor coherence length $\xi$. The second model applies to delocalized excitations along the wire's edges, which are the relevant low-energy excitations if $W \gtrsim \xi$. While our results were obtained for a simplified model of a p+ip superconductor, we expect that the general scaling properties presented here will remain present in a more realistic treatment of spin-orbit coupled superconductors.

The fermionic end states, which are the relevant low-energy excitations for $W \lesssim \xi$, appear in the topological as well as in the non-topological regimes, as soon as the wire has more than one channel, $W > \pi (k_{\rm F}^{2} - \xi^{-2})^{-1/2}$. For thin wires, $W \ll \xi$, the fermionic end states cluster around zero energy, see Fig. \ref{fig:Gap4}, so that the mere observation of a large local density of states near zero energy at the end of the wire is not a reliable signature of a Majorana bound state. Energy differences of order $\Delta \lambda_{\rm F} W/\xi^2$ must be resolved if one wants to separate the fermionic end states and the Majorana end states. This is a stricter requirement than for energy levels in vortex cores, which are spaced by a distance $\sim \Delta (\lambda_{\rm F}/\xi)$ \cite{Caroli1964}. The distance $\varepsilon_{\rm min}$ to the lowest-lying fermionic subgap excitation has a maximum for wire width $W \sim \xi$. The optimal value, $\varepsilon_{\rm min} \sim \Delta \lambda_{\rm F}/\xi$, has the same scaling as in a vortex core.

While the end states obscure the identification of a topological phase through a density-of-states measurement, the presence of localized fermionic states near the wire ends does not necessarily inhibit topological quantum computation based on the Majorana states \cite{Akhmerov2010,Moller2011}. 
In fact, the large gap $\sim \Delta $ to the delocalized bulk excitations, makes the narrow limit $W/\xi\ll1 $ favorable for topological protection. 
On the other hand, the lateral Majorana modes are delocalized, and their presence at low energy for $W \gtrsim \xi$ poses a restriction on the use of the Majorana end states for topological quantum computation.

It is important to point out that the end states we discussed here exist at the ends of multichannel superconducting wires. The situation is different at an interface between two regions in which the channel numbers changes by a small amount only. This is the relevant scenario if for semiconductor-based topological superconductors, where the change in channel number is induced by a small change in a gate voltage. In that case the number of fermionic subgap states localized at the interface is at most half the difference in channel number. In particular, if the channel number changes by one, the interface harbors a Majorana fermion, but no low-lying fermionic states.

We close with a remark on the effect of disorder. Even weak disorder is known to lead to the formation of subgap states localized somewhere in the bulk of the superconductor, with an algebraic density of states low energy \cite{Motrunich2001,Brouwer2011}. The effect of weak disorder on the low-energy end states, in contrast, is expected to be small, because the disorder potential has vanishing matrix elements between the basis states (\ref{eq:psin}) that are used for the construction of the effective end-state Hamiltonian (\ref{eq:Hmn}). 

We gratefully acknowledge discussions with M.~Duckheim, A.~Romito, V. ~Lahtinen, G.~M\"{o}ller, S.~Simon, Y.~Oreg and  F.~ von Oppen. This work is supported by the Alexander von Humboldt Foundation.

\appendix

\section{Appendix A}
\label{sect:AppA}
The bulk dispersion relation in two dimensions is given as $E_\bk = \sqrt{h_\bk^2  +|\xi_\bk|^2} $ with 
\bea
{h}_\bk = - \mu - 2 t (\cos k_x a + \cos k_ya ), \\
{\xi}_\bk = 2 \Delta (\sin k_x a+i \sin k_ya).
\eea
The ground state may be described by the BCS state \cite{Read2000}
\be 
\ket{\text{gs}} = \prod_{(\bk,-\bk)}  (u_\bk + v_\bk a^\dagger_\bk a^\dagger_{-\bk} )\ket{\text{vac}}. 
\label{eq:BCS}
\ee
where the product is over distinct $(\bk,-\bk)$ pairs and
\bea
\label{eq:uv}
u_\bk &=& \sqrt{1/2(1 + h_\bk/E_\bk)}, \\
v_\bk &=& -\sqrt{1/2(1 - h_\bk/ E_\bk)} e^{i \arg{\Delta_\bk}}.
\eea
This model has a topological phase transition at $|\mu|=4 t$.  The phase with $|\mu|> 4t$ corresponds to the strong pairing phase of Read and Green \cite{Read2000} and the Toric Code phase of the vortex free Kitaev honeycomb system \cite{Kitaev2006}. When $|\mu|< 4t$ the model is in the non-abelian weak pairing phase of Read and Green that supports bound Majorana fermions at vortex cores and along domain walls separating regions of topologically trivial and non-trivial phase. This phase is closely related with the Pfaffian filling fraction $\nu=5/2$ fractional quantum hall state and the gapped B-phase of the Kitaev honeycomb system \cite{Read2000,Kitaev2006}.

\section{Appendix B}
\label{sect:AppB}

We summarize the main results of the finite-width strip in the limit $W\ll\xi $, for the lattice model.  
Solving for the eigenstates of Eq. (3) in the main text for $\Delta_y=0 $ we find 
\be\label{eq:HS_size}
N= \mbox{int}\, \left[\frac{N_y+1}{\pi} \arccos\left( \frac{-2 \sqrt{t^2-|\Delta_x|^2} -\mu}{2t}\right)\right],
\ee
zero energy Majorana states with creation/annihilation operators $\gamma_{n,\pm} = e^{\mp i \pi/4} \sum_{i,j} u_{n,\pm}(ia,ja) (c_{i,j} \pm i c_{i,j}^{\dagger})$, $n=1,2,\ldots,N$, and
\begin{equation}
  u_{n,+}(x,y) = \sqrt{\frac{2}{N_y+1}}\Omega_n^{-1/2} r^{x/a} \sin\left(\frac{n \pi y}{N_y+1}\right)
  \sin(k_{nx} x).\ \
\end{equation}
Here 
\be
  r= \sqrt{\frac{t-|\Delta_x|}{t+|\Delta_x|}},
\ee
 $u_{n,-}(x,y) = u_{n,+}(L-x,y)$, and $\Omega_n $ is given by an analogous expression to $\Omega_{k_x} $ see Eq. \eqref{norm}. Accounting for the pairing $\Delta_y$ in degenerate perturbation theory then results in the effective $N \times N$ matrix Hamiltonian
\begin{eqnarray}
H_{nm} &=& \frac{4 i \Delta_y}{(N_y+1)} \frac{ \sin k_ma \sin k_na }{\cos k_ma -\cos k_na}\non\\
&\times& \frac{4\Delta_x^2 \sqrt{(b_m^2-4t^2)(b_n^2-4t^2)} }{\Delta_x^2 ((b_m + b_n)^2 -16t^2)  -(b_m - b_n)^2 t^2 },
\end{eqnarray}
where $b_n =\mu+2t\cos k_na $. In the continuum limit, one recovers Eq. (7) of the main text. 

\section{Appendix C}
\label{sect:AppC}

Here we review the calculation of the edge mode structure for the two-dimensional strip with periodic boundary conditions in the longitudinal ($x$) direction. If the boundaries at $y=0$ ($+$) and $y=W$ ($-$) are far apart, it is possible to exactly solve for the edge spectrum and recursively resolve the corresponding edge modes for arbitrary values of $k_x$.  It is well known, see for example \cite{Read2000,Tewari2007} that the modes on the well separated (upper or lower) edges are described by the Majorana ansatz
\be
\gamma_{k_x,\pm}^\dagger =   \sum_{m,n} u_{k_x,\pm}(n a) e^{i k_x ma}  ( c_{m,n}^\dagger  \pm c_{m,n})
\label{eq:Maj}
\ee
For the lattice model in question, the bottom (top) edge spectrum is given by $E=\pm 2 |\Delta_x| \sin(k_x a)$, see for example Refs. \cite{Mong2011,Zhou2011}.  Using this energy eigenvalue we can, on the bottom edge for example, calculate the unnormalized values of $u_{k_x,+}(n a)$ recursively through the transfer matrix as 
\be
\left[\begin{array}{c} u_{k_x,+}(na) \\ u_{k_x,+}((n+1)a) \end{array}  \right] =T_{k_x} \left[\begin{array}{c} u_{k_x,+}((n-1) a) \\ u_{k_x,+}(na) \end{array}  \right] 
\ee
with 
\be
T_{k_x}=\left[\begin{array}{cc} 0 & 1 \\ -C/A & -B/A  \end{array}  \right] 
\ee
and $A=t+|\Delta_y|$, $B=\mu+ 2 t \cos(k_xa)$, and $C=t-|\Delta_y|$. More generally we have
\be
\left[\begin{array}{c} u_{k_x,+}(na) \\ u_{k_x,+}((n+1)a) \end{array}\right] = [T_{k_x}]^n \left[\begin{array}{c} u_{k_x,+}(0) \\ u_{k_x,+}(a) \end{array}\right] 
\ee

The eigenvalues of the transfer matrix are $\lambda_\pm
=(-B\pm \sqrt{4AC-B^2})/2A$. 
The corresponding eigenvectors are 
\be
V_\pm=\frac{1}{d} \left[\begin{array}{cc} 1 & 1 \\ \lambda_+ & \lambda_- \end{array} \right]  
\ee
where $d=\sqrt{1+C/A}$. Note that these eigenvectors are not orthogonal. The inverse of the eigenvector matrix is 
\be
V^{-1}_\pm = \frac{d}{\lambda_-\lambda_+} \left[\begin{array}{cc} \lambda_- & -1 \\ -\lambda_+ & 1 \end{array} \right]  .
\ee
which allows one to write
\bea
[T_{k_x}]^n &=& V_\pm \left[\begin{array}{cc} \lambda^n_- & 0 \\ 0 & \lambda^n_+ \end{array} \right] V_\pm^{-1} \\ \non
&=& \frac{1}{\lm-\lp} \left[\begin{array}{cc} \lm \lp^n - \lm^n \lp & \lm^n-\lp^n \\ \lm \lp^{n+1} - \lm^{n+1} \lp &  \lm^{n+1}-\lp^{n+1} \end{array} \right].
\eea
For hard wall boundary conditions where $u_{k_x,+}(0) =0$ we would have 
\bea
\left[\begin{array}{c} u_{k_x,+}(na) \\ u_{k_x,+}((n+1)a) \end{array}\right] &=& T^n \left[\begin{array}{c} 0 \\ u_{k_x,+}(a) \end{array}\right] \\
&=&\frac{1}{\lm-\lp} \left[\begin{array}{c} \lm^n-\lp^n  \\ \lm^{n+1}-\lp^{n+1}  \end{array} \right] u_{k_x,+}(a),\non
\eea
and by setting $u_{k_x,+}(a)=1$ we find
\be
\label{eq:psil}
u_{k_x,+}(y)= \frac{\lm^n-\lp^n }{\lm-\lp} u_{k_x,+}(a) = \frac{r^{y/a} \sin{k_y y }}{\sqrt{\Omega_{k_x}}}.
\ee
Here 
\begin{eqnarray}
r&=& \sqrt{C/A} = \left(\frac{t-|\Delta_y|}{t+|\Delta_y|}\right)^{1/2} \approx e^{-a/\xi}\\
\nonumber 
k_y&=&\frac{1}{a}\arccos{\left\{\frac{-(\mu+2t\cos{k_xa})}{2\sqrt(t^2-\Delta^2)}\right\}}\\
&\approx& \sqrt{k_F^2-k_x^2-\xi^{-2}}.
\end{eqnarray}
The normalization of the wavefunction is given by $\sqrt{\Omega_{k_x}}$ where $\Omega_{k_x} = (s_1-s_2-s_3)/2$ with 
\bea
\label{norm}
s_1&=& \frac{2 (r^{2(N_y+1)}-1)}{r^2-1} \non \\
s_2&=& \frac{\exp(\phantom- 2 i (N_y+1)k_ya) r^{2(N_y+1)}-1}{(r^2) \exp(2 i k_ya)-1} \non \\
s_3&=& \frac{\exp(-2 i (N_y+1) k_ya) r^{2(N_y+1)}-1}{(r^2)\exp(-2 i k_ya)-1}.
\eea

The coupling between  Majorana modes at $y=0$ ($+$) and $y=W$ ($-$) is given by an effective 1-d, low-energy Hamiltonian, which in the Majorana basis $\gamma_{k_x,\pm} $ takes the form:
\be
H_{k_x}  = \left[\begin{array}{cc} \xi_{k_x} &h_{k_x}  \\ h_{k_x} & -\xi_{k_x} \end{array} \right] 
\label{eq:HMAJ}
\ee
where 
\bea
\xi_{k_x}  &=&   2\Delta \sin k_x a\xrightarrow{a\rightarrow 0} k_x\Delta/k_F\\
h_{k_x} & =& \delta \epsilon_m \langle u^{c}_{k_x,+} (y)|u^{c}_{k_x,-} (y)\rangle 
\eea
with
\bea\label{eq:d.E}
\nonumber
\delta \epsilon_m
&=&i\left[\mu+2t\cos(k_x a)
+2\sqrt{t^2-|\Delta|^2}\cos k_y a \right]\\
&\xrightarrow{a\rightarrow 0}&  -i \frac{\hbar v_{\rm F}}{2k_{\rm F}} [(k_y^{{\rm c}})^2 - k_y^2] 
\eea
and
\bea \label{eq.overlap}
 \langle u^{(b)}_{k_x} (y)|u^{(t)}_{k_x} (y)\rangle  &=&  (-1)^{m+1}r^{N_y+1}\frac{N_y+1}{2\Omega_{k_x}}\\
  &\approx& (-1)^{m+1}\frac{W}{2\xi}e^{-W/\xi}
\eea
The energy of (\ref{eq:HMAJ}) take the usual form $E_{k_x} = \sqrt{ h^2_{k_x} + |\xi_{k_x}|^2}$, and vanishes for $k_x=0 $ when
\be
k_y^{{\rm c}}=\arccos{\left\{\frac{-(\mu+2t)}{2\sqrt(t^2-\Delta^2)}\right\}} = \frac{\pi m_y}{N_y+1}.
\ee

\end{document}